\documentstyle[preprint,aps,floats]{revtex}
\tightenlines
\input psfig.tex 
\begin{document}
\def\ba{\begin{eqnarray}}
\def\ea{\end{eqnarray}}
\def\be{\begin{equation}}
\def\ee{\end{equation}}
\def\({\left(}
\def\){\right)}
\def\[{\left[}
\def\]{\right]}
\def\lagrange {{\cal L}}
\def\del {\nabla}
\def\d {\partial}
\def\Tr{{\rm Tr}}
\def\half{{1\over 2}}
\def\fourth{{1\over 8}}
\def\bibi{\bibitem}
\def\S{{\cal S}}
\def\H{{\cal H}}
\def\xx{\mbox{\boldmath $x$}}
\newcommand{\phpr} {\phi_0^{\prime}}
\newcommand{\gam}{\gamma_{ij}}
\newcommand{\sqgam}{\sqrt{\gamma}}
\newcommand{\dph}{\delta\phi}
\newcommand{\om} {\Omega}
\newcommand{\dom}{\delta^{(3)}\left(\Omega\right)}
\newcommand{\rar}{\rightarrow}
\newcommand{\Rar}{\Rightarrow}
\newcommand{\labeq}[1] {\label{eq:#1}}
\newcommand{\eqn}[1] {(\ref{eq:#1})}
\newcommand{\labfig}[1] {\label{fig:#1}}
\newcommand{\fig}[1] {\ref{fig:#1}}
\def\gsim{ \lower .75ex \hbox{$\sim$} \llap{\raise .27ex \hbox{$>$}} }
\def\lsim{ \lower .75ex \hbox{$\sim$} \llap{\raise .27ex \hbox{$<$}} }
\newcommand\bigdot[1] {\stackrel{\mbox{{\huge .}}}{#1}}
\newcommand\bigddot[1] {\stackrel{\mbox{{\huge ..}}}{#1}}
%\twocolumn[\hsize\textwidth\columnwidth\hsize\csname @twocolumnfalse\endcsname

\title{\bf An Observational Test of Quantum Cosmology}
\author{
Steven Gratton\thanks{email: S.T.Gratton@damtp.cam.ac.uk},Thomas
Hertog\thanks{Aspirant FWO-Vlaanderen;  email: T.Hertog@damtp.cam.ac.uk}
 and Neil
Turok\thanks{email: N.G.Turok@damtp.cam.ac.uk}}
\address{
DAMTP, Centre for Mathematical Sciences, Wilberforce Rd,
Cambridge, CB3 OWA, U.K.}
\date{\today}
\maketitle

\begin{abstract}
We compute the tensor CMB anisotropy power spectrum for singular
and non-singular instantons 
describing the beginning of an open universe
according to the Euclidean no boundary proposal. 
Singular
instantons occur generically, whereas 
non-singular instantons require more
contrived 
scalar field potentials. For the latter, we consider potentials
in which a sharp feature, either negative or positive, is added
to a gently sloping potential. In the first case 
one finds a nearly divergent 
contribution to the low multipole CMB anisotropy, in conflict
with the COBE observations. In the second case
the divergence is weaker, but matching the low multipoles 
forces the added feature to be large and narrow. 
For singular instantons, 
there is
a better match to the observations, without any such contrivance.
The distinction between singular
and nonsingular instantons disappears in the limit as the universe becomes
flat, but is still observable for densities as
high as 0.7 of the critical density.

\end{abstract}

\vskip .2in

\section{Introduction}

In the most common approach to inflationary theory one 
postulates a scalar field with a gently sloping potential, and 
assumes that  
for some reason the field was initially displaced
from the potential minimum. 
If the initial displacement is large, the field approaches 
a slowly rolling state in which the universe inflates. 
This state is an attractor, and in it
the system loses memory of the initial conditions. 
This scenario, which is certainly the simplest version of inflationary
theory, predicts that the universe 
should be flat to high accuracy today. It also predicts that
the initial state of the universe should be totally inaccessible
to observations today, since the scales most relevant to defining
the initial state were stretched by inflation to scales 
currently exponentially larger than the Hubble radius.
If future measurements confirm the universe is very nearly flat, 
then, assuming inflation is the explanation,  discussions of what came 
before inflation although interesting will remain strictly academic.

Current CMB observations are consistent with a flat universe, for example
the recent Boomerang measurement  \cite{melch} yields
$0.65<\Omega_{tot}<1.45$ at 95 per cent confidence. This lends support
to the hope that the simplest version of inflation, described above, 
 might be correct. 
However, significant space curvature is not yet excluded by the observations.
This paper is devoted to examining the observational consequences of
inflationary scenarios in which significant space curvature 
would exist today, and in which the initial conditions for
the open universe are actually visible in the microwave sky.
In an open universe, the curvature scale of the universe
on the surface of last scattering subtends an angular scale of
approximately $\sqrt{\Omega_0}$ radians, about 25 degrees for
$\Omega_0=0.3$. If we live in such a universe, cosmic
microwave sky observations can probe
the initial conditions for the inflating universe. 

Theories of open inflation were initially constructed from
scalar field potentials with false vacua, using instantons
known as Coleman-De Luccia instantons \cite{bgt}. These
can be interpreted as describing tunnelling from a prior
false vacuum inflationary state \cite{coldel,rubakov}, although
the relevant instantons only exist for rather special potentials. 
More recently, however,
it was realised that open inflation 
can occur far more generically 
through a class of singular, but finite action, instantons
\cite{HT}
which exist for essentially all gently sloping inflationary potentials. 
The regular instantons do have the virtue that the prediction
of $\Omega$ is unique in a given theory. For the Hawking-Turok instantons,
the most probable universe {\it a priori} is one with a very low
value of $\Omega$, but there are solutions for essentially all
values of $\Omega$ up to unity. In the absence of a better understanding
of how the actual value is determined, which may involve some
sortof anthropic considerations, we shall here simply treat the
value of $\Omega$ as a parameter to be adjusted to fit the universe 
we see. The pattern of density perturbations is then, for given
$\Omega$ and given scalar potential, uniquely predicted. 

In this paper we exhibit an interesting observable
difference between non-singular and singular instantons. 
We discuss a generic problem faced by non-singular instantons 
and show how it is alleviated in singular instantons. 

\section{Gravitational Instantons and Open Inflation}

Instantons are saddle point solutions of the Euclidean
path integral, and open inflationary instantons
may be naturally interpreted within the framework of
Euclidean quantum gravity and the 
no boundary proposal
\cite{HH}. 
The instantons 
provide a saddle point, which one can expand around to
compute the Euclidean  
path integral. Correlators of interest  
are
uniquely defined in the Euclidean region, and 
then analytically continued
into the Lorentzian universe.
We have recently carried this program through
to leading (quadratic) order for scalar and tensor 
perturbations
\cite{gt,tom}. (Related calculations 
were performed in a different approach in refs. 
\cite{garriganew,xavi}.) The well known problems 
of the non-positivity
of the Euclidean Einstein action and the non-renormalisability
of quantum gravity do not enter 
in these low order
calculations.

Until recently the class of known cosmological instantons 
was quite limited. Coleman and De Luccia discovered the first
examples when generalising 
the problem
of the decay of a false vacuum in scalar field theory 
to include gravity \cite{coldel}. 
In the limit of weak gravity
the decay is well understood and occurs 
via bubble nucleation.
In a localised region of space 
the scalar field quantum tunnels
through the barrier stabilising the false vacuum. The bubble
so formed expands at the speed of light and inside it the scalar field 
rolls down to the true vacuum.
In the presence of gravity, 
instantons only exist for scalar field potentials
with a sufficiently sharp false vacuum (as shown in Figure 1).
The reason is that 
the gravitational instanton has finite size, $\sim M_{\mathrm{Pl}}/
\sqrt{V}$ where $M_{\mathrm{Pl}}$ is the Planck mass and $V$ the 
potential energy density. For an instanton to exist,
in which the scalar field is not constant, 
the scale of variation of the field must be smaller 
than the instanton size. But this scale of variation is
determined by 
the second derivative of the
potential in the region of the barrier, $|V_{,\phi\phi}| \equiv M^2$.
The condition for existence of Coleman--De Luccia instantons 
is therefore that $M^2 >> V/M_{\mathrm{Pl}}^2$.

\begin{figure}
\centerline{\psfig{file=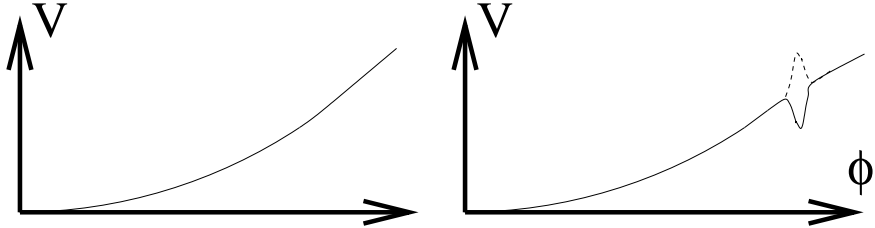,width=4.5in}}
\medskip
\caption{Inflationary potentials of the types being considered here.
The left is a smooth function, like $m^2 \phi^2 $,
$\lambda \phi^4$ or $e^{\epsilon \phi}$. In this theory 
only Hawking-Turok singular instantons exist. 
On the right are two potentials allowing 
 Coleman--De Luccia instantons. The solid line shows
a potential with a sharp false minimum added, the 
dashed line one with a sharp maximum. 
}
\labfig{pots}
\end{figure}

Coleman-De Luccia instantons may be used to describe the 
nucleation of bubbles in a false vacuum region of de Sitter space
\cite{coldel,rubakov}.
The interior of such bubbles then form infinite open universes
and with modest fine tuning of the distance $\Delta \phi$ over
which the field rolls during inflation, one can adjust
the value of $\Omega_{tot}$ to an interesting value less than unity 
today. But in order for the
 Coleman-De Luccia instanton to exist, the condition mentioned
above must be satisfied. 
Assume for example that
the potential is approximated by
${1\over 2} m^2 \phi^2$ in the neighbourhood of the true vacuum, where 
$m << M_{\mathrm{Pl}}$. For
$N$ efolds of inflationary expansion, 
one requires $\phi$ to roll for $ 2 \sqrt{N} M_{\mathrm{Pl}}$ where
$M_{\mathrm{Pl}}$ is the reduced Planck mass. The false vacuum has
to be at least 
this far from the true vacuum. But existence of
the Coleman--De Luccia solution requires
$M^2 >> 4 N m^2$, and for reasonable $N > 40$ (for
acceptable $\Omega$ today), the scale $M$ must be at least
an order of magnitude
larger than $m$. As we show later, yet another tuning is
required in order to suppress the large angle CMB anistropies. 

In open inflation,
it was assumed that the scalar field became stuck in the false vacuum,
leading to large amounts of inflation, in the course of which
the universe 
approached perfect de Sitter space. Bubbles would nucleate 
in this de Sitter space, as the field tunnelled through the
barrier between the false and true vacuum states. Each bubble
contains  an infinite, inflating open universe
\cite{coldel,bgt}.

Coleman--De Luccia instantons have a different interpretation
in the no boundary proposal. There they are viewed as
classical solutions describing the rounding off
of the Lorentzian universe on a compact Euclidean region.
The Euclidean path integral 
uniquely specifies the spectrum of fluctuations inside the
bubble without the need for additional assumptions 
regarding the
pre-bubble era. 
Indeed all 
calculations that have been performed to date have in effect 
used the instanton background to define the pre-bubble
era. This makes the calculations identical to 
those  performed in the no boundary interpretation.
To that extent 
one can say that all predictions of open 
inflation are really predictions of the no boundary
proposal. The pre-bubble inflating universe appears to be a redundant 
theoretical construction.

The Coleman--De Luccia instantons are interesting because they
provide a calculable scenario for open inflation. As mentioned
the potentials needed to obtain such instantons are necessarily
contrived (Figure 1). However
Hawking and one of us recently showed that
a class of singular but
finite action Euclidean instantons
exists for almost  every gently sloping inflationary potential
\cite{HT}.  We have computed the spectrum
of fluctuations about such singular instantons and 
found that in spite of the singularity the 
correlators are uniquely defined, just as
in the 
Coleman--De Luccia case \cite{gt,tom}. Interesting differences arise
because the singularity imposes Dirichlet boundary
conditions on the perturbation modes. We pointed out that
the observational effect of this difference 
is likely to be most pronounced in the 
tensor spectrum and this is what we discuss here. 
We shall show that the
part of parameter space for Coleman--De Luccia theories in which
the bubble size is much smaller than the De Sitter radius,
so that the tunnelling is very similar to that in flat spacetime, 
is 
ruled out.

\section{Fluctuations about Singular and Non-Singular Instantons}

As we shall see, Coleman--De Luccia
instantons with potentials as shown in Figure 1 
generically produce a very large amplitude of long wavelength
tensor modes. Let us make clear at the outset however
that
this constraint cannot be used to rule out all such models. 
The tensor perturbations are
governed by the height of the inflationary potential,
and can be adjusted independently of the scalar perturbations
by flattening the potential.
For example potentials of the form used in hybrid inflation,
with a flat plateau followed by a
sharp drop produce acceptable scalar perturbations but
almost no tensor component. Another way of suppressing 
the low
multipoles is to
further tune the potential so that it is 
steep around the values of $\phi$ where the bubble nucleates
\cite{LindeSas}. In the light of our discussion above, this
constitutes a third fine tuning, needed to make such
models work. 

In this paper we compare the predictions of singular 
and Coleman-De Luccia
instantons with potentials of the form 
shown in Figure 1. In the singular case we assume a simple
monomial potential like $\phi^2$ or $\phi^4$, and in the
Coleman-De Luccia case we superpose
a sharp negative false vacuum. We show 
that unless $\Omega_{tot}$ is rather close to unity today 
the Coleman-De Luccia examples
are generally ruled out because they predict unacceptable
large angle anisotropies in the microwave sky. 
In contrast the singular instantons which
occur generically in gently sloping inflationary potentials
appear more compatible with the observations.

In previous work we have derived the two-point correlators of the 
scalar \cite{gt} and tensor \cite{tom} metric perturbations 
in open 
inflationary universes associated with both classes
of Euclidean cosmological instantons.
All perturbations are determined from
correlators of the gauge-invariant Newtonian potential
$\Psi_{N}^{\ }$ and the transverse traceless tensor perturbation $t_{ij}$,
which may be computed 
directly from the 
path integral.

To first order in $\hbar$ the 
Euclidean correlators are specified by a Gaussian integral
\cite{gt,tom}. 
For both regular Coleman--De Luccia instantons and 
singular instantons the result is unique. In the 
latter case 
the 
singularity enforces Dirichlet boundary conditions. 
The Euclidean two-point correlators are 
 analytically 
continued into the Lorentzian region where they describe the quantum 
mechanical vacuum fluctuations of the various fields
in the state described by the no boundary 
proposal initial conditions. 

In the present work we shall calculate
the temperature fluctuations on the 
microwave sky
from the Lorentzian two-point correlators.
The key observable difference between the two types of 
instantons occurs for wavelengths of 
order the curvature scale. 
Since the long-wavelength continuum in the scalar power 
spectrum vanishes linearly with wavenumber $p$,
the differences are small there. 
Likewise 
the bound state of the scalar perturbation potential,
producing long range correlations beyond the
curvature scale in the open universe, 
is known to have a very minor effect on
the CMB anisotropy \cite{garrev}.
However the 
spectrum of primordial gravity waves has for
the regular instantons a near divergence at small $p$
and therefore provides a better opportunity
for a distinguishing test.
 
The result for the symmetrised two-point
correlator tensor metric perturbation about either 
Hawking--Turok or Coleman--De Luccia instantons
is \cite{tom}
\begin{eqnarray}\label{lorcorf}
\langle \{ t_{ij}(x),t_{i'j'}(x')\}\rangle & = &
2\kappa \Re
\int_0^\infty \frac{dp}{p}\left(\coth p\pi
g_{p}(\tau )g_{-p}(\tau ')+ r_p
\frac{g_{p}(\tau)g_{p}(\tau ')}{\sinh p\pi }\right)
{W_{\ iji'j'}^{L (p)}(\chi) \over a(\tau) a(\tau')}
\end{eqnarray}
where $\kappa= 8\pi G$, and length units are chosen so that the spatial
curvature scale is unity. 
In this formula $\tau$ is the conformal time as defined in~\cite{gt} and $\chi$ the 
comoving radial coordinate. The
bitensor $W_{\ iji'j'}^{L(p)}(\chi)$ 
is the sum of normalised rank-two tensor harmonics with
eigenvalue $\lambda_{p} =-(p^2+3)$ of the Laplacian on $H^3$
\cite{tom}.
The eigenmodes $g_{p}(\tau)$ are solutions of the \emph{Lorentzian} tensor
perturbation equation
\begin{equation}
\left(-\frac{d^2}{d\tau ^2} +\frac{a''}{a} -1\right)g_{p}(\tau)
=p^2g_{p}(\tau)
\end{equation}
normalised to  obey $g_{p}(\tau) \rightarrow e^{-ip\tau}$ as 
$\tau \rightarrow -\infty$.

First note the potential $p^{-2}$ divergence in the integrand
due to the $1/(p$sinh$p \pi)$ in the first term. This
divergence, as we shall now argue, is cancelled by the second term. 
The second term involves 
$r_p$ which is a reflection amplitude computed in the Euclidean region. 
In conformal coordinates
the metric takes the form
$b^2(X) (dX^2 +d \Omega_3^2)$. For singular instantons 
we have $0<X <\infty$ where the singularity is located at
$X=0$. 
For regular instantons we have $-\infty <X <\infty$.
In both cases the perturbations obey a Schr\"{o}dinger-like equation 
with potential $U(X)\equiv {b''(X) \over b}-1$. 
This potential diverges to $+\infty$ at $X=0$ in the
singular case, but is finite everywhere in the regular case. 
In fact in the latter case it is close to a reflectionless potential
$-2$ sech$^2 X$. 
The quantity $r_p$ is in both cases the 
reflection amplitude for  waves incident from $X=+\infty$.
For singular instantons it is by unitarity a phase
but for non-singular instantons it is a complex number of modulus
less than unity, and it is small at high $p$. 
Both reflection amplitudes 
tend to minus one as $p \rightarrow 0$ because 
long-wavelength modes are completely reflected, hereby yielding
an infrared finite correlator.
However, since the non-singular Coleman--De Luccia instantons are much
closer to the perfect $S^4$ non-reflecting solution, we expect
the reflection amplitude to tend to
$-1$ at much lower $p$ than in the 
singular Hawking--Turok case. 

The region of low $p$ in the the tensor spectrum is what is known
in the literature as 
the bubble wall fluctuation spectrum \cite{garrev}. When the
de Sitter symmetry is only weakly broken, with a scalar field
present, there is a near divergence
in this spectrum. These 
long-wavelength tensor perturbations give a substantial 
contribution to the CMB anisotropies.
From the discussion above, we expect a larger contribution to the large angle
microwave anisotropies
for regular instantons. In other words, the mild breaking of
de Sitter invariance in non-singular models allows for large 
long-wavelength fluctuations about the background solution. On the contrary,
in singular models the deviation from an $O(5)$ instanton is drastic, 
and 
the singularity keeps such fluctuations small.
 
\section{Non-singular ``Thin-wall'' Instantons}

The scalar field equation in the Euclidean region reads
\ba
(b^3 \phi_{,\sigma})_{,\sigma}=b^3 V_{,\phi},
\labeq{sfe}
\ea
where $\sigma$ is the proper radial distance $(d\sigma= b dX)$. 
Following \cite{bgt} we 
consider the case where the potential is given by superimposing a sharp
negative ``bump'' of amplitude $-\Delta V$ centred about $\phi_f$ onto a
smooth monotonically 
increasing function of $\phi$.  On a non-singular instanton, the scalar field
rolls in the upside down potential from $\phi_0$, gaining kinetic 
energy until it hits the ``bump'' and rapidly decelerates to an almost
standstill near $\phi_f$.  Effectively all of the kinetic energy of
the field is converted to potential energy and any
damping is negligible.  The
field then remains approximately constant as the 
scale factor $b$ turns round and vanishes as $\(\sigma_m-\sigma\)$.
$\phi_0$ is fixed by the form of the potential and the requirement
of regularity. 
 This
generally implies that the scalar field must have reached the ``bump''
well before the scale factor turns round.  We therefore take
$b\approx\sigma$ in equation~\eqn{sfe}, and approximating $V_{,\phi}$
as $V_{,\phi_0} \equiv V_{,\phi}\(\phi_0\)$, we have
\ba
(\sigma^3 \phi_{,\sigma})_{,\sigma}\approx \sigma^3 V_{,\phi_0}.
\ea
We have $\phi_{,\sigma}=0$ at the regular pole,
so we may solve to find
\ba
\phi\approx \phi_0 + {1 \over 8} \ V_{,\phi_0} \sigma^2.
\ea
If the field approaches the ``bump'' at $\sigma_b$, then its kinetic
energy just before hitting the ``bump'' is
\ba
\half \phi_{,\sigma}^2\approx \frac{1}{32} V_{,\phi_0}^2
\sigma_b^2 \approx {1\over 4}  V_{,\phi_0} \(\phi_f-\phi_0\)
\ea
and we may equate this to $\Delta V$.

As we shall discuss shortly, it is useful to rewrite the 
Schr\"{o}dinger equation in the 
Euclidean region in a form where it involves the potential
$\overline{U}= \frac{\kappa}{2}\phi'^2$ where prime denotes derivative with
respect to conformal coordinate $X$. The strength of the potential is
then
\ba
C\equiv\int\frac{\kappa}{2}\phi'^2 dX = \int \frac{\kappa}{2} b
\phi_{,\sigma}^2 d\sigma \approx \frac{\kappa}{2}
\frac{V_{,\phi_0}^2 \sigma_b^4}{64}.
\labeq{ceq}
\ea

If we take the smooth part of the potential to be of the form $\lambda
\phi^n$, we may introduce the quantities $N\equiv
\frac{\kappa\phi_0^2}{2n}$ and $H^2 \equiv \frac{\kappa}{3}
V\(\phi_0\)$.  $N$ is the slow roll approximation to 
the number of inflationary efoldings in the open universe.  $H$ is the
slow roll Hubble parameter, with $b\approx \frac{1}{H} \sin
H\sigma$.  Then we can write
\ba
C = \frac{9 n \(H \sigma_b\)^4}{256 N}.
\ea
We will see below that for regular instantons the quantity 
$C$ provides an infrared cutoff for the amplitude of the bubble wall
fluctuations.
From the condition that the scalar field must have reached the bump
well before the scale factor turns round, using $b\approx \frac{1}{H} \sin
H\sigma$ we see that $H\sigma_b$ must certainly be less than $\pi
\over 2$.  As a concrete example, if we take $n=2$ and $N=50$, this yields
$C < 0.01$. Generically, in the regime where the bubble radius is 
much smaller than the radius of the de Sitter space, $C$ will be very much
smaller than this, since the formula involves the fourth power
of the size of the bubble, $\sigma_b$.

\section{Non-singular ``Thick-wall'' Instantons}

We now consider instantons associated with potentials with
a sharp {\it positive} feature as shown by the dashed curve in Figure 1.
In
this case, the scalar field motion is confined within the region of the
feature over the instanton, and does
not probe the smooth part of the potential at all.  Unlike the
thin-wall case discussed above, the scalar field varies significantly
over the whole instanton, and not just over a localised region 
of it.  The starting value of the scalar field is
tuned so that the field reaches the peak of the feature at
approximately the same time as the scale factor rolls over.  
If indeed the potential is exactly symmetrical about the peak over the
region probed by the instanton, these two events occur at
exactly the same moment.  
The
change in sign of the slope of the potential may then be able to balance
the antidamping, bringing the scalar field to a halt as the scale
factor again tends to zero, giving us a non-singular solution.  For
this to occur, the feature must be sufficiently sharp.  This can be
achieved using differentiable functions with large curvature at the
peak. We can model this by introducing a kink. 
We model the potential in the vicinity of the feature at say $\phi_*$ as
$V-V_{,\phi} \left|\phi-\phi_*\right|$, with $V$ and $V_{,\phi}$ constant and
positive.  
Then we approximate $b$ as $1/H \sin H\sigma$, with
$H^2=\frac{\kappa}{3} V$, assuming that $V$ dominates over gradient energy
in the field.  In this approximation
$\phi$ reaches $\phi_*$ at $\sigma=\pi/2H$, and $\phi_\sigma$ is
odd about $\sigma=\pi/2H$.
So in order to calculate
$C= \int \frac{\kappa}{2} b
\phi_{,\sigma}^2 d\sigma$, we need only work out $\phi_{,\sigma}$ up
to $\sigma=\pi/2H$ and multiply by two. 
From the scalar field equation we have
\ba
\phi_{,\sigma} & = & \frac{V_{,\phi}}{\sin^3 H\sigma}  \int_0^{\sigma}
\sin^3 H\sigma d\sigma \nonumber \\
& = & \frac{V_{,\phi} \(\cos^3 H\sigma - 3\cos H\sigma+2\)}{3 H \sin^3
H\sigma}
\labeq{phidot} 
\ea
and so
\ba
C &=& 2\times \frac{\kappa}{2} \frac{V_{,\phi}^2}{9 H^2} \int_0^{\pi/2H}
\frac{1}{H \sin^5 H\sigma}
\(\cos^3 H\sigma - 3\cos H\sigma+2\)^2 d\sigma \nonumber \\
&=& \frac{5}{4\kappa} \(\frac{V_{,\phi}}{V}\)^2. \labeq{cthick}
\ea

We can also integrate Eq.~\eqn{phidot} to find that $\Delta \phi \equiv 
\phi_*-\phi(0)= 1/2
(1+2 \ln 2) V_{,\phi} / (\kappa V)  \approx 1.19  V_{,\phi} / (\kappa V)$.  
Inserting into \eqn{cthick} we see that $C$ can be expressed 
two ways, either as $C \approx \Delta \phi V_{,\phi}/ V \equiv
\Delta V/V$, or as $C \approx
0.8 \Delta \phi^2/M_{Pl}^2$, where $M_{Pl}^2\equiv \kappa^{-1}$.
We have checked that the above
expressions match the numerically calculated values quite closely
up to $C \sim 1$.
In order to get a value of $C$ close to unity, 
one requires a large feature in the potential - i.e.
a large change in $V$ to occur over a range of $\phi$ 
which is at least of order 
unity in (reduced) Planck units. The calculations shown
below exclude small values of $C$, corresponding in the 
thick wall case to 
small positive features on the potential.

\section{Euclidean Reflection Amplitudes and  Modes}

The primordial gravity wave spectrum is given by
equation (\ref{lorcorf}). In terms of the proper distance 
$\sigma$ we used in the previous section,  we shall 
fix the integration constant involved in defining the
conformal coordinate $X$ by setting 
$X=\int_{\sigma}^{\sigma_t}\frac{d\sigma'}{b\(\sigma'\)}$.
For
non-singular instantons we  follow references~\cite{gt}
and~\cite{tom} and define 
$\sigma_t$ to be that value of $\sigma$ for which $b$ is
maximum. For singular instantons it is taken instead to be the value of 
$\sigma$ at the singularity. 

For singular instantons, the singularity acts to impose Dirichlet 
boundary conditions in the Euclidean region. The only allowed mode function 
at fixed $p$ is given by 
$\psi_p\rar a_p e^{ipX} +
a_{-p} e^{-ipX}$ as 
$X\rar\infty$.
In the non-singular case, we have left and right moving modes. 
The left mover is 
$g_p^{\mathrm{left}}\(X\)\rar e^{-ipX},\,$ as $X\rar -\infty$ and
$g_p^{\mathrm{left}}\(X\)\rar c_p 
e^{ipX}+d_p e^{-ipX}$ as $X\rar +\infty$.  These mode functions 
satisfy the
differential equation
\ba
\(-\frac{d^2}{dX^2}+U(X)\)g_p\(X\)=p^2
g_p\(X\). 
\ea
which 
has a trivial bound state solution $b(X)$ with $p^2=-1$.  
This corresponds to a constant shift in the metric perturbation
which is pure gauge. It is very convenient to project this out,
since its presence means that there is an extra phase shift of $\pi$
produced by the potential even at very low $p$. The projection is
simple. Rather 
than $g$ one considers 
$\bar{g}\equiv b\(\frac{g}{b}\)'$ which is clearly zero 
for the bound state \cite{garriganew}.  
This variable also satisfies a Schr\"{o}dinger equation 
\ba
\(-\frac{d^2}{dX^2}+\bar{U}(X)\)\bar{g}_p\(X\)=p^2 \bar{g}_p\(X\)
\ea
where $\bar{U}$ is the positive-definite quantity $\frac{\kappa}{2}
\phi'^2$ mentioned in the previous section.
 We define $\bar{g}_p\(X\)$ in an identical fashion.
From the constancy of the Wronskian and using  $b\sim e^{-|X|}$
at the regular poles, one finds 
$r_{p}=\frac{1-ip}{1+ip} \bar{r_{p}}$.
For singular instantons the reflection amplitude $r_{p}$
is given by the phase $\frac{a_{p}}{a_{-p}}$, and in the
non-singular case it equals $\frac{c_{p}}{d_{p}}$.
It
is straightforward to 
calculate $\bar{r_{p}}$ numerically for any 
background instanton of interest. 

 For the non-singular instantons
considered in the previous sections $\bar{U}(X)$ is sharply 
peaked around a value of $X$, $X_b$ say.
We can then make a very good analytic 
approximation to ${\bar{c}_p \over \bar{d}_p}$ as follows.  We replace
$\bar{U}(X)$ by the delta function potential $C \delta(X-X_b)$ of
equivalent strength, with $C$ as defined in Eq.~(\ref{eq:ceq}) or
Eq.~(\eqn{cthick}) as required.   
 We can then solve
analytically for 
$\bar{g}_p^{\mathrm{left}}\(X\)$ and find
\ba
{\bar{c}_p \over \bar{d}_p}=
-\frac{\(1+\frac{2ip}{C}\)e^{-2ipX_b}}{1+\frac{4p^2}{C^2}}.
\labeq{ratio}
\ea
We approximate $X_b$ as follows.  In the ``thin-wall'' case $X_b$
corresponds to $\sigma=\sigma_b$.  Then 
 with $b\approx \frac{1}{H} \sin
H\sigma$,
\ba
X_b \approx \int_{\sigma_b}^{\sigma_t} \frac{H}{\sin H\sigma} d\sigma
\approx -\ln \tan \frac{H \sigma_b}{2}
\ea
where we have used $H \sigma_t\approx \frac{\pi}{2}$.  In the
``thick-wall'' case $\bar{U}\(X\)$ is simply peaked around $\sigma_t$
and so $X_b \approx 0$.   

\section{Tensor CMB Anisotropy in Open Inflation}

The Euclidean no boundary  path integral 
allows us to compute correlation functions of any observable. 
If these correlations are well approximated by 
a classical statistical distribution, as macroscopic
observables such as the microwave anisotropies are, we can regard the 
predictions as being characterised by the classical distribution. 
Our observed universe is one member of this classical
ensemble.  
To compare different theories with
regard to an observation carried out in our universe, we compute 
how likely the given observation is according to each theory.

We consider the microwave background anisotropy generated
by primordial fluctuations, expanded in spherical
harmonics
${\delta T}/T = \sum a_{lm} Y_{lm}\(\theta,\phi\).$
The $a_{lm}$'s obey 
$a_{lm}^*=a_{l-m}$ and we have $2l+1$ real observable
quantities for each $l$.  Rotational invariance implies
that the 
 $2l+1$ quantities are independently distributed with
zero mean and common variance $C^{\mathrm{th}}_l$.  Neglecting higher-order
effects, their probability 
distributions are Gaussian.  For a given $l$ we average
over the squares of the $2l+1$ observable quantities in our universe
to determine $C^{\mathrm{obs}}_l$.  Then for a given theory of this type,
$C^{\mathrm{obs}}_l / C^{\mathrm{th}}_l$ is $\chi^2$-distributed over the ensemble
of universes with $2l+1$ degrees of freedom.  $C^{\mathrm{obs}}_l$ itself is
gamma-distributed with probability density function
$f\(C^{\mathrm{obs}}_l;\(l+\half\) / C^{\mathrm{th}}_l,l+\half\)$~\cite{rpp}.  If
$C^{\mathrm{obs}}_l$ is 
greater than the median value of $C^{\mathrm{th}}_l$, then the fraction of
universes with $C_l$ less likely than $C^{\mathrm{obs}}_l$ is given by $2
\Gamma\(l+\half,\(l+\half\)C^{\mathrm{obs}}_l / C^{\mathrm{th}}_l \) /
\Gamma\(l+\half\)$.  Similarly, if $C^{\mathrm{obs}}_l$ is
less than the median value of $C^{\mathrm{th}}_l$, then the fraction of
universes with $C_l$ less likely than $C^{\mathrm{obs}}_l$ is given by $2-2
\Gamma\(l+\half,\(l+\half\)C^{\mathrm{obs}}_l / C^{\mathrm{th}}_l \) /
\Gamma\(l+\half\)$.

We need to obtain the $C^{\mathrm{th}}_l$'s for the different theories we are
interested in. Using the usual Sachs-Wolfe formula \cite{SW} this is given 
in terms of our symmetrised tensor correlator (\ref{lorcorf}) as
%We have already computed the two-point tensor correlator (\ref{lorcorf})
%in an open universe \cite{tom}.
%The contribution of gravitational waves to the
%CMB anisotropy is given by the integral in the Sachs-Wolfe 
%formula \cite{SW},
%\begin{eqnarray}\label{temp}
%\frac{\delta T_{SW}^{\ }}{T}(\theta,\phi)& = &
%-\frac{1}{2}\int_{\tau_{e}}^{\tau_{0}}d\tau
%t_{\chi \chi,\tau}^{\ }(\tau,\chi,\theta,\phi)|_{\chi = \tau_{0} -\tau}
%\end{eqnarray}
%where $\tau_{0}$ and $\tau_{e}$ are respectively the observing and
%last scattering time for the photons and $\chi$ is the 
%comoving radial coordinate.
%The anisotropy is conventionally
%characterised by the two-point angular correlation function
%\begin{equation}\label{ang}
%C(\gamma) =\left\langle \frac{\delta T}{T}(0)\frac{\delta T}{T}(\gamma)
%\right\rangle =
%\sum_{l=2}^{\infty} \frac{2l+1}{4\pi} C_{l} P_{l}(\cos \gamma)
%,\end{equation}
%where $\gamma$ is the angle between two points on the celestial sphere.
%Hence, by inserting the Sachs-Wolfe integral (\ref{temp}) into (\ref{ang})
%%and after writing the bitensor $W_{iji'j'}^{(p)L}$ back in terms of its
%defining eigenmodes $q_{ij}^{plm}=Q_{ij}^{pl}Y_{lm}$ 
%on $H^3$, one obtains a direct relation between the 
%two-point tensor correlator (\ref{lorcorf}) and the multipole moments 
%$C_{l}^{\mathrm{th}}$:
\begin{eqnarray}\label{multi}
C_{l}^{\mathrm{th}} = 
\kappa \Re \int_{0}^{+\infty}\frac{dp}{2p}
\int^{\tau_{\mathrm{now}}}_{\tau_{\mathrm{lss}}}d\tau
\int_{\tau_{\mathrm{lss}}}^{\tau_{\mathrm{now}}}d\tau'
&& \left( \coth p\pi
\left[\dot\Phi_{p}^{L}(\tau )\dot\Phi_{-p}^{L}(\tau')\right] \right. \nonumber\\
&& \quad \left.+  \frac{1}{\sinh p\pi}
\left[r_{p}\dot\Phi_{p}^{L}(\tau )
\dot\Phi_{p}(\tau')\right]\right)Q^{pl}_{\chi\chi}
Q^{pl}_{\chi'\chi'}.
\end{eqnarray}

The primordial tensor power spectrum at the end of inflation defines inital 
conditions for the Sachs--Wolfe integral.
To compute the multipole moments we use CMBFAST \cite{sz},
which evolves the mode
functions from the surface of last scattering at $\tau_{\mathrm{lss}}$
up to the present
time $\tau_{\mathrm{now}}$, given the initial power spectrum. 
Modifications were required to improve the resolution at low
wavenumbers, necessary for the accurate evaluation of the 
the low $l$ multipoles.
We then
combine the tensor component in the correct ratio \cite{ratio}
with the standard scale invariant 
scalar spectrum of perturbations
in order to obtain the total $C^{\mathrm{th}}_l$ to compare
with experiment.   
To extract the primordial tensor power
spectrum from equation (\ref{lorcorf}),
we first construct approximate solutions for the eigenmodes
$g_p^L\(\tau\)=\Phi^{L}_{p}(\tau) a(\tau)$.  
In the 
inflationary phase of the open 
universe the mode functions closely follow
perfect de Sitter evolution in which they tend to 
a constant after the physical wavelength has been stretched outside
the Hubble radius. Hence, to determine the amplitude and phase of this 
constant we approximate $a\(t\)$ as $\frac{1}{H} \sinh Ht$ until
$\Omega$ is close to one, and introduce the associated conformal
coordinate  
\ba
\eta \equiv - \int_t^{\infty} \frac{H}{\sinh Ht} dt = \ln \tanh
\frac{Ht}{2},
\ea
$\eta \rar -\infty$ being the start of inflation, and $\eta \rar 0$ as
the universe continues formally to inflate without end. $\tau-\eta$ is
a finite constant during inflation whilst this 
approximation for $a(t)$ is a good one.     
The approximate Lorentzian tensor
perturbation equation is then
\begin{equation}
\left(-\frac{d^2}{d\eta ^2} +\frac{2}{\cosh ^2 \eta} \right)f^L_{p}(\eta)
=p^2f^L_{p}(\eta)
\end{equation}
and the solution in which $f^L_{p}(\eta) \rightarrow e^{-ip\eta}$ as 
$\eta \rightarrow -\infty$ is
\ba
f^L_p(\eta)=\frac{ip+\coth \eta}{ip-1} e^{-i p \eta}.
\ea
At a given value of $t$ then, with corresponding $\tau$ and $\eta$, we
have $g^L_p\(\tau\) \approx e^{-ip(\tau-\eta)} f^L_p(\eta)$.  So 
dividing by $a$ and taking the late-time limit we see that
\ba
\Phi_p\(\tau_0\) \approx  -H \frac{e^{-ip\tau_0}}{ip-1}.
\labeq{approxphi}
\ea
Here $\tau_0$ is the conformal time as defined in~\cite{gt} at the end
of inflation.  This can be calculated numerically and is $O(1)$ for
singular instantons and $O(0.01)$ for ``thin-wall'' non-singular
instantons.  From 
equation (\ref{lorcorf}) the primordial 
tensor power spectrum $P_{\mathrm{T}} (p)$ at the end of inflation is
\ba
2\kappa \Re \frac{1}{p} \left(\coth p\pi
\Phi_{p}(\tau_0
)\Phi_{-p}(\tau_0)+\frac{1-ip}{1+ip}\bar {r_{p}} 
\frac{\Phi_{p}(\tau_0)\Phi_{p}(\tau_0)}{\sinh p\pi }\right).
\labeq{powerend}
\ea
For singular
instantons $\bar{r_{p}}=\bar{a}_p / \bar{a}_{-p}$ is a phase factor and can be written as
$e^{2i\bar{\theta}_p}$.  So the tensor power spectrum $P^{\mathrm{S}}_{\mathrm{T}} (p)$ for
singular instantons is
\ba
P^{\mathrm{S}}_{\mathrm{T}} 
(p)=\frac{2\kappa H^2}{p\(1+p^2\)} \(\tanh \frac{p\pi}{2} +
\frac{1}{\sinh p\pi}\(1+\frac{1}{1+p^2} \cos 2\(\bar{\theta}_p-p\tau_0\) \)\)
\labeq{powerht}
\ea
in this approximation.  For a given potential one evaluates
$\bar{\theta}_p$ numerically and obtains an empirical fit.
In the long-wavelength limit $\bar \theta_{p} \rightarrow -\pi/2$
so the power spectrum is infrared finite. Actually, it turns out
that the CMB power spectrum predicted by singular instantons
differs only a little from the one with 
a perfect reflecting potential in which
the ratio $a_{p}/a_{-p}$ is replaced by $-1$ for all $p$.

%For example, for
%a $\frac{1}{2} m^2 \phi^2$ potential, with $N=50$,
%\ba
%\theta_p \approx  \(-.25+\frac{.75}{1+.75p+.3p^2}\)\pi.
%\ea

For non-singular instantons, we have $\bar {r_{p}}=\bar{c}_p/\bar{d}_{p}$
in~\eqn{powerend}.  Using our approximations for this in the previous
section, we obtain
\ba
P^{\mathrm{NS}}_{\mathrm{T}} (p) = \frac{2\kappa H^2}{p\(1+p^2\)} \( \tanh \frac{p\pi}{2} +
\frac{1}{\sinh p\pi} \( 1-\frac{\cos 2p\(X_b+\tau_0\)+\frac{2p}{C} \sin 
2p\(X_b+\tau_0\)}{1+\frac{4p^2}{C^2}} \) \).
\labeq{powercd}
\ea

Equations (\ref{eq:powerht}) and (\ref{eq:powercd}) 
define the initial conditions for the numerical computation of the
Sachs--Wolfe integral for the different models.
A Taylor expansion of the second term around $p=0$ shows that in the
``thin-wall'' case, as 
speculated earlier \cite{tom}, for typical values of $C$
the regime $c_{p}/d_{p}
\rightarrow -1$ sets in at much lower $p$ than in the 
singular Hawking--Turok case. One can see from (\ref{multi}) that this
leads to a larger contribution to the large angle microwave anisotropies 
for regular ``thin-wall'' instantons.
In the next section we discuss to what extent this characteristic feature 
of false vacua models allows one to observationally distinguish them from 
singular open inflation models.

\hfill\break 
\begin{figure}
\centerline{\psfig{file=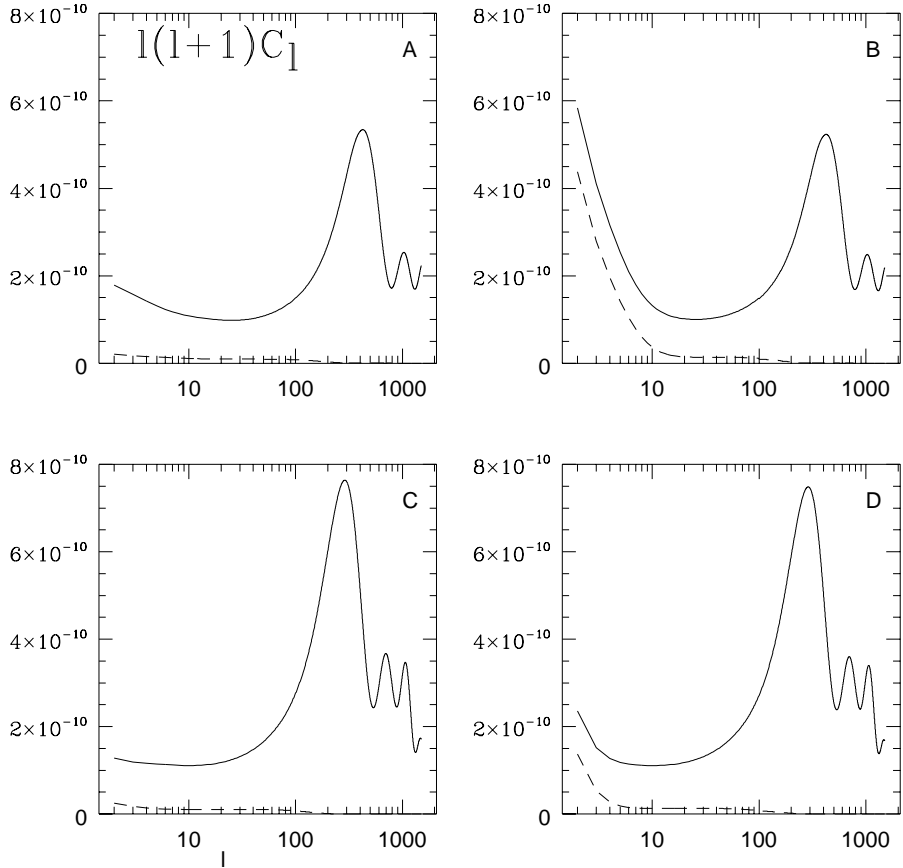,width=5.0in}}
\caption{Cosmic microwave sky predictions of different cosmological
instantons. The upper two panels show predictions for an open universe,
with $\Omega_{tot}= 0.3$ and no cosmological constant, for 
Hawking--Turok (panel A) and Coleman--De Luccia instantons (panel B).
The former is
for an $\frac{1}{2}m^2\phi^2$-potential, the latter for a model
where a false vacuum has been added (see text). The lower
panels compare the 
Hawking--Turok (C) and Coleman--De Luccia (D) theories for 
an $\Omega_{matter}=0.3$,  $\Omega_{Lambda}=0.4$, $\Omega_{tot}=0.7$ 
cosmology. The difference at low $l$ is still marked. 
These 
results are for a cold dark matter dominated universe
with $\Omega_{CDM}=0.25$, baryon density 
$\Omega_{B}=0.05$, and Hubble constant $h=0.65$.}
\labfig{graphs}
\end{figure}

\section{Numerical Results}

In Figure~\fig{graphs} we compare the CMB anisotropy
power spectra for singular and nonsingular instantons, in open
universes with $\Omega_{tot}=0.3$ and $0.7$ respectively.
For the nonsingular instantons there is a large contribution from
the tensor component, shown by the dashed line. The amplitude
of the large angle contribution is governed by the parameter $C$
discussed above. We argued above that on general grounds $C$ has to be
smaller than 0.01 for `thin-wall' instantons. For `thick-wall' instantons
$C$ can be larger if the feature on the potential is large. 
For singular instantons there is no such parameter to vary.
We
have chosen $C=0.025$ for the nonsingular instantons 
which is certainly conservative for the thin wall case. 
The divergence
at low $l$ would be even more pronounced in the allowed regime.

These calculations show that even if the 
curvature of the universe today is quite modest, one nevertheless
sets strong constraints on the form of the inflationary potential
and on the nature of the primordial instanton. 
As emphasised above,
$C$ decreases as the fourth power of the size of the
bubble wall thickness - if the latter is much smaller 
than the Hubble radius of the de Sitter space, $C$ is
much smaller than 0.01. Since the amplitude of
the correlator diverges as $\int dp p^{-2}$, the amplitude of
the quadrupole diverges roughly as $C^{-1}$. 

In the panels shown the result for
the gravity wave spectrum explained above has been combined 
with the usual scalar spectrum of perturbations appropriate for
an open universe \cite{bgt}.
The ratio of tensor to scalar contributons 
is a function of cosmological and model parameters. However,
for medium multipoles, $(l \sim 30 )$, the ratio approaches
its well known flat space value \cite{ratio}. This value then
fixes the relative normalisation of scalar versus tensor
anisotropy for all multipoles. 
For a $\lambda\phi^n$ inflaton potential the flat space
ratio $R_{\mathrm{fl}}=0.05 n$. Therefore, the
higher the value of $n$ is, the more important the contribution from the
bubble wall tensor fluctuations.  
In the plots shown we have taken the ratio to be that for $n=2$.
This yields a quadrupole ratio 
$R_2 \approx 0.13$ in a singular model and $R_2 \approx 0.57$ in a regular
model. Higher values for $n$ would allow us to exclude the
nonsingular models more strongly.

For both singular and non-singular instantons 
we see the rise in $C_l^{\mathrm{th}}$ at low $l$, characteristic of an open
universe.  We compare these different models to the COBE DMR data as
follows.  First of all have to set the overall normalization of each
model.  We do this with the RADPACK software 
~\cite{lloyd,bond2}. Using
the DMR data alone, we find the 
normalization which maximizes the likelihood for each model.
We then compare likelihoods amongst the different models.  The
relative likelihoods
are as follows:

\begin{center}    
\begin{tabular}{ccccccccc}
Singular & : & Singular with $\Lambda$ & : & Non-singular & : &
Non-singular with $\Lambda$ & : & Flat spectrum\\
28 & : & 76 & : & 1 & : & 22 & : & 97
\end{tabular}
\end{center}

where the flat spectrum is one with constant $l\(l+1\)
C_l$'s, shown for comparison. In this, Bayesian, approach,
the nonsingular instantons for an open universe with $\Omega_{tot}=0.3$
are strongly disfavoured. 

Having done the likelihood analysis above, we now take an
hypothesis-testing approach, using the probabilities given in
Section VII above. The strongest constraint on the models
comes from the quadrupole, and we focus on that here. 
Unfortunately the true sky quadrupole is not yet known, 
and the literature contains various estimates of it.
We have therefore assumed a range of values taken from various references.
It is to be hoped that the MAP experiment will accurately determine
the actual value.

The model dependence of the low $l$ $C_l$'s 
suggests that in order to
quantify the difference between the models we should compare the
correctly-normalized 
predicted quadrupole moments with the measured quadrupole moment.  In
the table below we show the percentage of universes in the ensemble
associated with a given theory with a measured quadrupole more extreme
than that seen.  We compare the results from singular and
non-singular instantons with the 
best fit flat  
spectrum for comparison.  
We have done this for a selection of groups'
estimates for the 
observed quadrupole~\cite{bunn,bond,gor}.  We have converted all
measured values to the dimensionless quantity $l\(l+1\) C_l / 2\pi$,
dividing by $\(2.73\, \mathrm{K}\)^2$ where necessary, to match the
output of CMBFAST.  The result from~\cite{bunn} is effectively a
direct measurement of the quadrupole, albeit with a systematic error due to
the galactic cut.  The other results are harder to interpret,
having been obtained using maximum
likelihood techniques with highly 
non-Gaussian likelihood functions for the quadrupole~\cite{bond}. This
means that the quoted values below should have large skewed error bars. 
We also show what a measurement of a larger quadrupole
(that from the best fit flat spectrum) would tell us
for illustration.   

\begin{center}
\begin{tabular}{|c||c|c|c|c|c|} \hline
~Measured value~ & ~Singular~ & ~Singular~ & ~Non-singular~&
~Non-singular~ & ~Flat spectrum~ \\
&$\Omega_{\Lambda}=0$&$\Omega_{\Lambda}=0.4$&
$\Omega_{\Lambda}=0$&$\Omega_{\Lambda}=0.4$& \\

$3 C_2 /\pi$&$1.7 \times 10^{-10}$&$1.2 \times 10^{-10}$&$4.0 \times
10^{-10}$&$2.3 \times 10^{-10}$&$1.0\times 10^{-10}$~\cite{bond} \\
\hline 
$0.11\times 10^{-10}~\cite{bunn}$   & 0.56\% & 1.2\% & 0.071\% &  0.27\%
&  1.9\% \\ \hline 
$0.20\times 10^{-10}~\cite{bond}$   &  2.3\% & 4.9\% & 0.31\% & 
1.1\%  & 7.0\% \\ \hline 
$0.37 \times 10^{-10}~\cite{gor}$  &  9.0\% & 18\% & 1.3\%    & 4.6\%
&  25\% \\ \hline 
$1.0 \times 10^{-10}~\cite{bond}$  &  61\% & 96\% & 13\% &  37\%   &
83\% \\ \hline 
\end{tabular}
\end{center}

We note that in general the probability is 
is several times larger for the singular case as compared to the 
the non-singular case. 
Both 
models are easier to rule out at a given confidence level than the
flat spectrum.  Note that even this model is
ruled out at the 98\% level if the result of~\cite{bunn} is taken at face
value!  It should be remembered that for our non-singular
``thin-wall'' model, we
have assumed a value of $C$ even larger than the  extreme best-case.  In
situations where 
the theoretical quadrupole is much larger than the measured one, the
probability scales as $\(C_2^{\mathrm{meas}}/C_2^{\mathrm{th}}\) \sim
C^{5/2}$ for the non-singular models.  The ``thin-wall''
nonsingular instantons appear to be strongly ruled out by the 
observed quadrupole, even if $\Omega_{tot}$ is as large as 0.7.

%From~\cite{max} $l(l+1)C_l^{obs}$ for $l=2$ is $9.7 \times 10^{-12}$.
%This is much lower than the theoretical prediction in both cases,
%indicating that neither fit the data too well.  With $C_2^{NS}=2.37
%\times 10^{-10}$ and $C_2^{S}=1.76 \times 10^{-10}$, we can compute
%the fraction of universes in each ensemble with $C_2$ more extreme
%than the observed value as discussed above.  For the singular case
%this fraction is $4.2 \times 10^{-3}$, whereas for the non-singular
%case it is $2.0 \times 10^{-3}$.  While it is disappointing that both
%numbers are so low, we note that the singular one is more than
%double the non-singular `best case' one.  In other words, to create an
%open universe an
%$\frac{1}{2}m^2\phi^2$ inflaton potential is favoured by COBE above a similar
%potential with a false vacuum superimposed.
%Decreasing $C$ - by altering the position of the
%false vacuum for instance -
%makes the non-singular prediction worse. 

\section{Conclusion}

We have computed the 
 tensor CMB anisotropy power spectrum for a class of
singular and non-singular instantons. We showed that this 
provides a way to 
observationally distinguish different versions of open inflation.
The ``thin-wall'' 
false vacuum models 
generate larger
fluctuations on large angular scales, distinguishing them 
from singular models.
Using the COBE data, we have found that
this characteristic feature strongly disfavours 
``thin-wall'' Coleman--De Luccia
instantons 
relative to the singular Hawking-Turok models. 
Non-singular ``thick-wall''  Coleman--De Luccia
instantons are still viable, but only if the false vacuum
feature 
in the scalar potential is large. In this case
the predictions depend strongly
on the detailed parameters describing the feature and 
the models are hence somewhat unattractive.
These calculations 
have therefore enabled us to further constrain the form of the 
inflaton potential in open inflation.

\medskip
\centerline{\bf Acknowledgements}

This work was supported by a PPARC (UK) rolling grant, an EPSRC
studentship and a PPARC studentship. We thank L. Knox
for providing the RADPACK sofware used for the likelihood
analysis above.  We thank M. Bucher, J. Garriga, X. Montes, V. Rubakov,
M. Sasaki, T. Tanaka and other participants in the Isaac Newton
Institute programme {\it Structure Formation in the Universe }
for very helpful discussions.

\end{document}